# Simulation-based Safety Assurance for an AVP System incorporating Learning-Enabled Components[*]

Hasan Esen (h.esen@eu.denso.com)
Brian Hsuan-Cheng Liao (h.liao@eu.denso.com)

DENSO AUTOMOTIVE Deutschland GmbH
Freisingerstr. 21-23, D-85386, Eching Germany

## 1. Abstract

There have been major developments in Automated Driving (AD) and Driving Assist Systems (ADAS) in recent years. However, their safety assurance, thus methodologies for testing, verification and validation AD/ADAS safety-critical applications remain as one the main challenges. Inevitably AI also penetrates into AD/ADAS applications, such as object detection. Despite important benefits, adoption of such learned-enabled components and systems in safety-critical scenarios causes that conventional testing approaches (e.g., distance-based testing in automotive) quickly become infeasible. Similarly, safety engineering approaches usually assume model-based components and do not handle learning-enabled ones well. The authors have participated in the public-funded project FOCETA[1], and developed an Automated Valet Parking (AVP) use case. As the nature of the baseline implementation is imperfect, it offers a space for continuous improvement based on modelling, verification, validation, and monitoring techniques. In this publication, we explain the simulation-based development platform that is designed to verify and validate safety-critical learning-enabled systems in continuous engineering loops.

## 2. Introduction

Automated Driving Assist Systems (ADAS), such as lane-keeping assist or adaptive cruise control have been well established in the automotive industry. In parallel, rapid developments in AI technologies lead ADAS to evolve into higher level automations, and Automated Driving (AD) systems became an ambitious and fascinating target of automotive companies. SAE J3016 standard[2] defines 6 levels of automation, of which last three (Lvl3 – Lvl 5) incorporates automated driving functions. From Lvl 4 on, the AD system does not require a human interaction and can perform the driving task on its own. Therefore, it is also common to refer Lvl4+ systems as Highly Automated Driving systems (HAD).

Testing, verification and validation (V&V) of HADs is one of the main challenges in the automotive industry. Wachenheim et al. published that to guarantee the safe operation of a HAD system (without fatal accident) at least 6.62 billion test kilometers must be driven [WAC-2016]. In fact, one can hear from automotive industry members even higher numbers recently. The main message to derive is that conventional testing approaches (e.g., distance-based testing in automotive) is infeasible to give safety assurances to HADs.

Consequently, automotive industry in collaboration with academia have been researching on HAD safety assurance, V&V methodologies and tools for the past 10 years. It is possible to distinguish 2 pioneering projects in this domain. The first one is Pegasus project[3]. Its' main objective was to define a standardized procedure for the testing and experimenting of automated vehicle systems in simulation, on test stands and in real environments. This to be accompanied by a continuous and flexible tool chain to safeguard the automated driving. The second project is Enable-S3[4], which had a multi-domain approach to accelerate application of highly automated and autonomous systems in automotive, as well as in aerospace, rail, maritime and health care domains; based on virtual testing, verification and coverage-oriented test selection methods.

---

[*]  This project has received funding from the European Union's Horizon 2020 research and innovation programme under grant agreement No 956123 - FOCETA.
[1] FOCETA - FOundations for Continuous Engineering of Trustworthy Autonomy, foceta-project.eu, Horizon 2020 Framework Grant Agreement N° 956123
[2] J3016_202104: Taxonomy and Definitions for Terms Related to Driving Automation Systems for On-Road Motor Vehicles - SAE International

As a partner of Enable-S3 project, we led an Automated Valet Parking Use Case, and developed a baseline safety verification platform for this purpose. In the frame of a follow-up project, FOCETA, this work was extended with the methologies / tools for verification and validation of learning-enabled components and systems; as well as with continuous engineering approaches to give incremental safety assurances.

In this paper, first the AVP use case (Sec 3) and our V&V architecture and platform (Sec 4) are introduced. Following that we will explain how our safety case was build and how it has been verified using our V&V platforms (Sec 5). Sec 6 is conclusion, including exploitation of the results and future research outlook.

## 3. Automated Valet Parking (AVP) Use Case

In an AVP scenario, the driver drops off his/her car in front of a parking area or drop-off zone and authorizes an automated system to take control to autonomously guide the vehicle through the parking area and into a designated parking bay. Similarly, the vehicle is coordinated back to a pick-up zone after a respective user request [ESE-2020]. Figure 1 illustrates a snapshot of a potential AVP scenario.

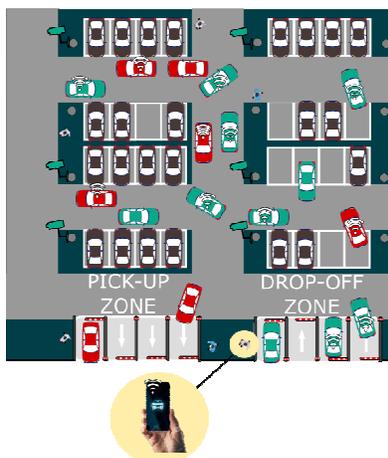

*Figure 1. Automated Valet Parking*

AVP is expected to be among the first commercially available automated driving (AD) systems with a Level 4 classification, i.e., a fully automated driving system without a required user interaction. Supporting reason for this fact is that in low-speed AD, such as AVP, the vehicle can be stopped in a short amount of time, which represents a safe state. Additionally, for parking environments, a clear definition of the operational design domain (ODD) can be described. For example, the parking area can be geometrically defined in high detail and delimited from other areas such as public roads.

---

[3] PEGASUS project (https://www.pegasusprojekt.de/en/), funded by the German Federal Ministry for Economic Affairs and Energy (BMWi), 2016-2019
[4] Enable-S3project (https://cordis.europa.eu/project/id/692455) , funded by H2020 ECSEL-IA - ECSEL Innovation Action, Grant agreement: 692455, 2016-2019.

Nevertheless, parking areas can present unstructured environments due to many factors. For instance, walking pedestrians whose moving pattens are hard to model, occluding building structures, and a high variability of possible scenarios. Finding solutions to these challenges requires suitable and scalable safety-by-design methods as well as structured and rigorous verification and validation (V&V) procedures.

## 4. Verification & Validation Platform Architecture for AVP System

The holistic approach of FOCETA project foresees 4 main pillars to assess trustworthiness of learning-enbled autonomous systems: (1) Modelling and Simulation, (2) V&V, (3) Monitoring, and (4) Continuous Engineering [BEN-2023]. The baseline V&V platform, developed in Enable-S3 project, was extended following FOCETA project holistic approach as given in Figure 2.

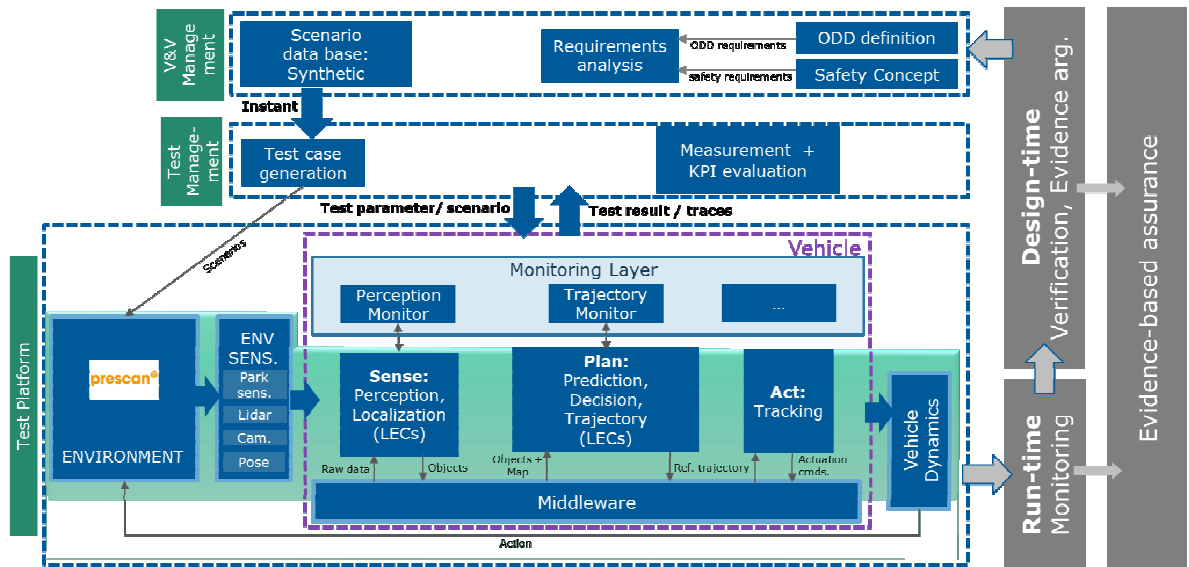

*Figure 2. Platform architecture for AVP use case*

We consider three layers (from top to bottom):
- V&V Management: The system description and definition such as the ODD, the safety concept, and requirements.
- Test Management: Test cases are derived and prepared in the correct format to be executed. It acts as a design-time monitoring layer, which observes and evaluates components and/or overall AVP system.
- Testing Platform: The virtual simulation environment (System Under Test) implementation of the AVP functionality. Additionally, this layer includes run-time monitoring functions.

FOCETA project partners have developed several design-time monitors and run-time monitors, which are integrated into given AVP architecture. Such monitors indicate safety-critical deficiencies, which is used to update relevant component, requirement, definitions, and thus closes the continuous engineering loop.

Following subsections highlight important features of each architectural layer in more detail.

### 4.1. V&V Management: Requirements, Scenarios and ODD

Fig. 3 shows more insight regarding the main elements of the V&V Management layer. It concerns describing a safety concept for verification and validation of the AVP system. After describing functional AVP scenarios, they are parameterized based on Operational Design Domain (ODD) definition. This leads to creation of a scenario database, which contain all initial concrete scenarios to be used at lower layers to define concrete, critical test cases. Safety concept also contains a set of safety requirements, which are analyzed by requirement monitors and eventually updated if monitor detects missing and/or contradicting requirements. Additionally, requirements management block contains tools to convert the requirements from native language into formal languages such as Signal Temporal Logic (STL).

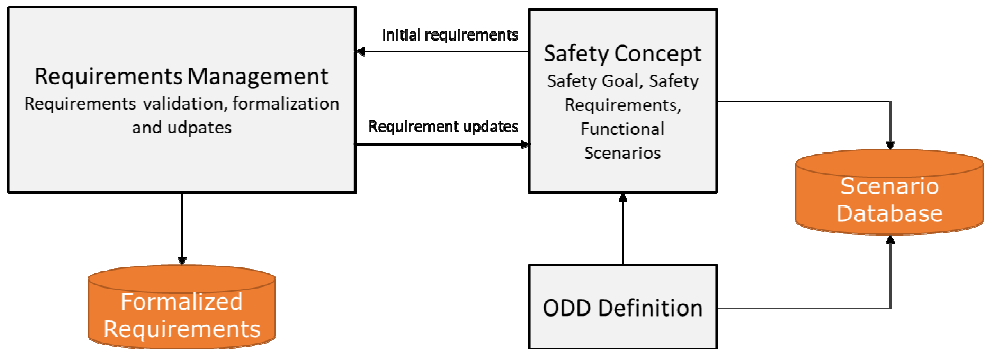

*Figure 3. Verification & Validation management layer*

### 4.2. Test Management: Design-time Monitoring
Test Management layer's objectives are 2 folded:
(1) Monitor components during their design-time and identify potential faulty behaviors in advance. Accordingly, improve the component design.
(2) Monitor the overall SUT simulation results and lead the testing towards critical scenarios using optimal sampling strategies.

An overview of the test management layer is given in Fig 4. Several component and system level design-time monitors have developed by FOCETA partners. In Sec 5, the perception testing component FuzzOD, developed by the authors, will be explained in more detail. For other design-time monitors mentioned in the figures, refer to: OpenSBT [SOR-2023]; RAM [ZHA-2021]; PyTorchALFI [2023].

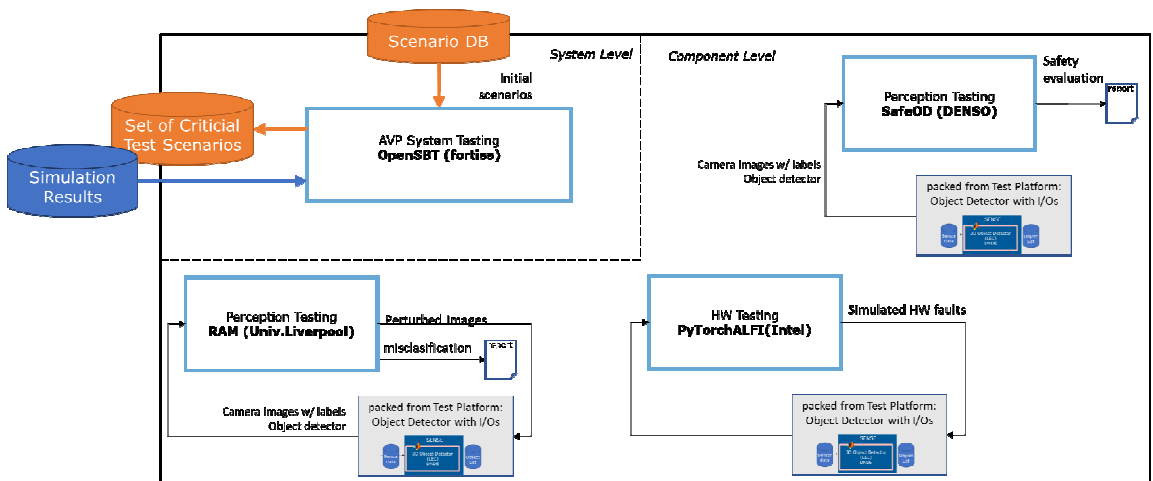

*Figure 4. Test Management layer*

### 4.3. Test Platform
Test platform has 2 fundamental sub-layers: System Under Test (SUT) and Run-time Monitoring.

**SUT** is the simulation environment that incorporates all the necessary component models and middleware to execute a full-fledged AVP system as shown in Fig 5. The environment model includes the parking area map, ego vehicle and static / dynamic obstacles. One can obtain from the the environment the ground truth information revealing several facts such as the position and type of the obstacle. This information is necessary specially to test whether the perception sensors and functions are working properly. Therefore, in addition to the ground truth, it is possible to integrate environment sensors and include the effect of the sensors in the perception. We have developed a 3D object detector based on machine learning principles. This component converts the sensor data (camera input images) into a detected object list. "Planning" action start with geometrical path planning, based on given parking environment map, ground truth information on obstacles and destination point. This is followed by a static trajectory planner, which sets the target velocity and steering of the vehicle on the given path. Finally, a dynamic planner refers to object list and adjust the velocity and steering targets accordingly. In our case, we used an automated emergency braking (AEB) system as dynamic planner. The target values are then tracked by Trajectory Tracker, which is also a learning-enabled component developed by FOCETA partner Siemens [VOO-2023]. Finally, the requested signals are sent to vehicle dynamics model, and the ego vehicle in the virtual environment is controlled.

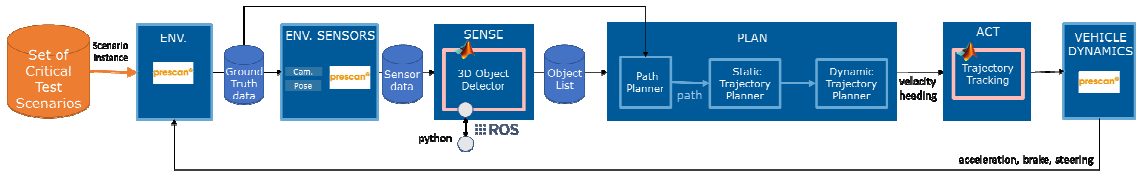

*Figure 5. System Under Test sub-layer*

SUT receives from test management layer a set of critical test scenarios, which are executed and tested against system requirents. Tool chain is based on SIEMENS SimCenter prescan and Matlab / Simulink. Object detector also uses a ROS bridge to link to required python libraries.

**Run-time monitoring** sub-layer observes component or system behavior against requireiments during run-time. Considering the ever-changing world, a component or a system can hardly be fully assured with design-time measures only, thus, run-time monitors are inevitable. Such monitors on the one hand enable to record abnormalities during run-time, which could then be used to improve the component / system insufficiencies at next design step. On the other hand, such monitors can enforce safe actions during run-time.

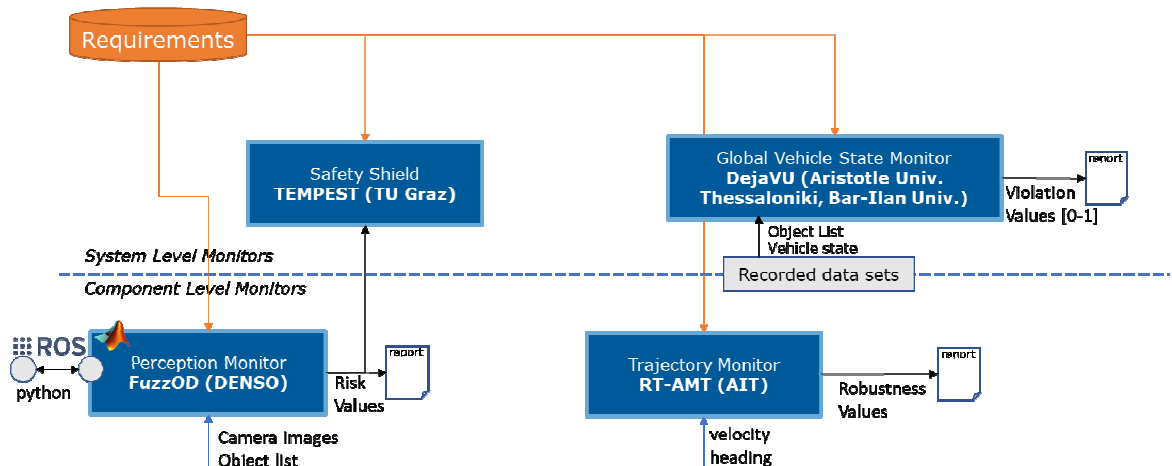

*Figure 6. Run-time Monitoring sub-layer*

Multiple run-time monitors have been developed in FOCETA project and integrated into the AVP baseline system as given in Fig 6. In Sec 5, we will explain the perception monitor FuzzOD in more detail. Readers interested in other monitors can refer to: Safety Shield TEMPEST [PRA-2021]; RT-AMT [NIC-2020], DejaVU [HAV-2018].

## 5. Testing AVP System

In this section, we will showcase a concrete example how the different architectural layers and blocks defined in Sec 4 are practically linked together to give a safety assurance for AVP system. The analysis focuses on the object detector component.

### 5.1. Setting up the Requirements, ODD and Scenarios

**Functional scenario description:**
In AVP use case, our main safety concern is a potential collision with pedestrians. Therefore, our functional scenario target is the collision avoidance against pedestrians. The scenario is sketched in Fig. 7. Two pedestrians walk out from the parking spots, with potential occlusion by the closer pedestrian on the further one (from ego vehicle's perspective). The second pedestrian steps into the ego vehicle's driveway while it is driving in the parking lot toward the designated parking spot.

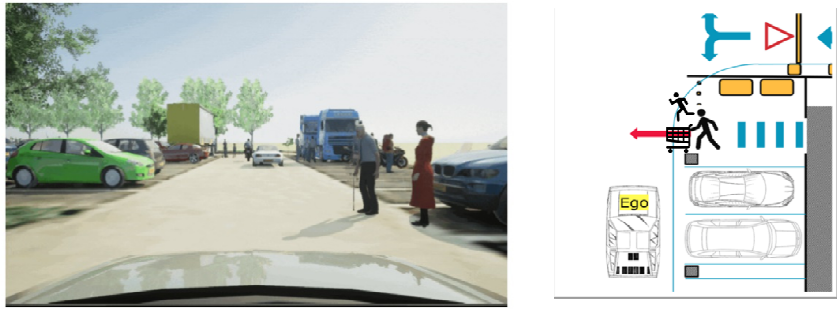

Figure 7. AVP functional scenario example

**Safety Goal:**
Within the ODD, the AVP system has no collision against pedestrians, unless the ego vehicle has a velocity of zero (i.e., in the safe state).

**Operational Design Domain (ODD):**
The ODD is defined by the intended operation area and limited by the laws and regulations applying to the domain. Our automated system is intended to work in defined mixed traffic parking areas with a designated drop-off zone. The speed limit for parking areas is set at 10 km/h in many European countries. The expected speed of other traffic participants can be faster, and this behaviour must be detected and risks must be mitigated by the AVP system capabilities. The static environment is regulated by varying laws, e.g., in the state of Bavaria of Germany, there is a law for parking garage design[5]. A shortened list of AVP-ODD elements is given in Table 1.

Apart from the given examples, several other ODD elements regarding environment illumination, weather condition, expected actors in the evrionment and the density of the pedestrians are defined.

---

[5] Verordnung über den Bau und Betrieb von Garagen sowie über die Zahl der notwendigen Stellplätze, Abbr.: Garagen- und Stellplatzverordnung – GaStell (https://www.gesetze-bayern.de/Content/Document/BayGaV)

| No | Item | Description |
|---|---|---|
| ODD-01 | Entrance/Exit | Dedicated entering and exiting lanes must exist. |
| ODD-02 | Parking lots | Parking lots must be at least 5 meters in length and 2.3 meters in width. |
| ODD-03 | Driving lanes | Driving lanes must be at least 3 meters (if one-way traffic) or 5 meters (otherwise) in width. |
| ODD-04 | Driving lanes | Driving lanes must be free of snow, ice or any road safety impairing conditions. |
| ODD-05 | Ego vehicle | The ego vehicle's velocity is bounded within (-1 m/s, 2.8 m/s). |
| ODD-06 | Ego vehicle | The ego vehicle's acceleration is bounded within (-7 m/ss, 2 m/ss). |
| … | … | … |

*Table 1. Example AVP – ODD elements*

**Safety Requirements:**
Based on-the functional scenario and ODD definition and considering the issue of SOTIF (safety of the intended functionality) described in the standard ISO PAS 21448 [SOT-2022], we derived several component-level and safety-level safety requirements for the AVP system. Below in Table 2 a few selected requirement examples are given.

| Req. No. | Item | Description |
|---|---|---|
| UC-AVP-01 | ObjDetect | Within D_safety meters from the ego vehicle, the object detection component shall identify pedestrians in their correct position, subject to an error tolerance of E_detect meters, with sufficiently large dimension and correct velocity. |
| UC-AVP-02 | SelfLocal | The position of the ego vehicle shall be measured, subject to an error tolerance of E_local meters. |
| UC-AVP-03 | PathPlan | The planner shall calculate a reference path that keeps a distance D_safety to obstacles. |
| UC-AVP-04 | TrajTrack | The actual path shall not diverge from the reference path for more than E_ meters. |
| UC-AVP-05 | EmBrake | The brake shall be fully applied to generate a deceleration of at least -7 m/ss if a pedestrian is detected within D_safety meters and with a time to collision less than T_safety seconds from the ego vehicle. |
| UC-AVP-06 | Sense+Plan+Act | The computation time of a sense-plan-act cycle should be maximally T seconds. |
| … | … | … |

*Table 2. Example AVP component and system requirements*

**Scenario Database:**
Based on the functional scenario and the ODD parameter ranges given above, several concretes test scenarios are generated, and stored in the scenario database. Those scenarios are passed to critical scenario identification block in test management layer. Generated critical scenarios are the final test instances to be fed to SUT layer.

### 5.2. Critical Test Case Generation

In scenario-based testing (SBT) of safety-crticial systems, test inputs are generated by tailoring meta-heuristic optimization approaches. That means, for each executed test case, a fitness value is assigned and heuristic-specific operations on test inputs are performed to generate test inputs with better fitness values, so that finally failure-revealing test cases are found. For the execution of a test case with the SUT, preferably a realistic simulation environment is used. When several fitness functions have to be defined because several objectives have to be optimized such as the distance to actors or the distance of the car to the center line, then the testing problem is modeled as a multi-objective optimization problem.

In FOCETA, we have utilized OpenSBT tool, developed by our project partner fortiss [SOR-2023]. It is a modular and flexible, open-source framework that allows the application of search-based testing to automated driving systems to identify failure-revealing test cases. OpenSBT tackles challenges such as the integration of the simulation environment, of the preferred search techniques, of search configurations based on the tester requirements (e.g., specification of the fitness function, scenario), as well as the automated output and analysis of testing results.

### 5.3. Object Detection

Object detection refers to the function of simultaneously classifying and localizing objects within a single-frame input. Such a functionality is extremetly safety-criticial for autonomous driving (AD) applications. Missing a close-by pedestrian in front of the driving path could possibly result in fatal cases.

We developed a method, SafeOD, to formally verify the safety of 3D object detectors based on spatial properties. Specifically, we formulate a safety requirement and mathematically formalize it into a straightforward safety metric by combining the Intersection-over-Ground-Truth (IoGT) measure and a distance ratio. Subsequently, we formulate a safety-aware loss function. We propose to use the safety metric to model the relation between the prediction and the ground-truth when evaluating the safety of 3D object detectors, and we also use the safety-aware loss function to improve their safety [LIA-2023].

We developed an object detector based on FCOS3D [WAN-2021] and applied our safety-aware loss function for its fine-tuning. To train and test the model, synthetic data sets are collected utilizing Siemens Simcenter Prescan simulator. Data sets include camera images from parking environment (e.g., humans, cars, trucks, …), ground-truth annotations (bounding-boxes), and metadata information (e.g., calibration information among global, ego and camera coordinates). The data are recorded and maintained using the KITTI dataset format[6].

We compared the results of using our safety-oriented loss function (FCOS_s) against using the state-of-the-art accuracy-based Smooth-L1 loss (FCOS_a) for a fine-tuning of 10 epochs. Results are depicted in Figure 8.

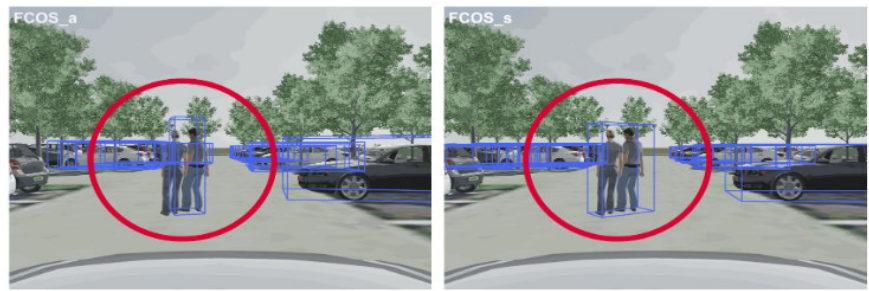

(a) Qualitative results.

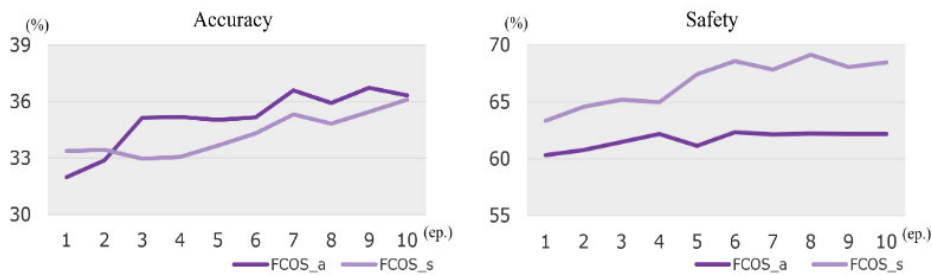

(b) Quantitative evaluation.

*Figure 8. Safety improvement results of the 3D object detector SafeOD in the AVP use case [LIA-2023].*

---

[6] The KITTI Vision Benchmark Suite.

8 / 12

Qualitatively, the FCOS_s model produces a better bounding box to contain the entirety of the pedestrians; quantitatively, it is observed that FCOS_s delivers a similar number of accurate predictions to FCOS_a yet reduces many unsafe cases. More details regarding this work can be found in [LIA-2023].

### 5.4. Perception Monitoring

We designed an anomaly-detecting monitor, FuzzOD, based on a fuzzy inference system to safeguard a given object detector and estimate the safety risk of its predictions during run-time.

The basis of our monitor is a state-of-the-art depth and motion estimator, rigidmask [YAN-2021], which has been shown effective in providing reliable cues for downstream computer vision tasks. Our core idea is to compute the consistency of the cues and the object detector's predictions and map it to safety risk via fuzzy logic. Since there is a lack of "ground truth" during run-time and our visual cues are not always precise, we believe such fuzzy logic-based reliability assessment is an essential safety feature.

Our monitor has been applied in the AVP use case. Results indicate that if we place an unkown object (e.g., animals), the object detector usually fails to recognize the object, since such unknows objects were not a part of the training data set. However, our FuzzOD monitor starts to raise the risk level and passes this signal to the Safety Shield [PRA-2021] to stop the car.

### 5.5. Continuous engineering

Sections 5.1-5.4 explains how to define appropriate test scenarios and accordingly verify the functionality of a safety-critical component, such as an object detector in an AVP system. One important last step is to update regularly overall system design process (from requirements to component design) utilizing the outputs generated by design-time and run-time monitors, i.e., closing continuous engineering (CE) loops.

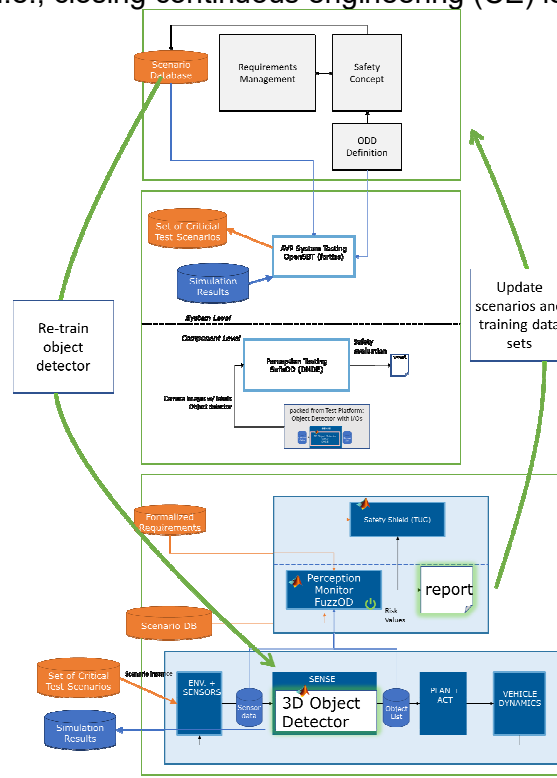

*Figure 9. The continuous engineering loop to update the learning-enabled object detector component.*

A good example for CE updates can be derived from the section 5.4. The object detector has been missing to detect animals in the parking lot, and however this situation was observed and reported by the FuzzOD monitor. Using this information, we can collect new data sets including animals in the scene, and re-train our object detector. In return this delta improvement would allow us also to detect animal in the next generation product. This CE loop is shown in the figure 9.

In addition to this run-time monitoring-based system repairs, we closed the CE loop in two other ways:
(1) Ontology-based requirement analyses and requirement updates: The requirements management tools such as RFT [MOK-2022] can perform continuous requirement analyses to detect missing or contradicting requirements. Accordingly, we can only remain in the highest layer (i.e., V&V Management) in our architecture and perform off-line updates on system requirements and specifications.

(2) Simulation-based test case generation, specification mining, and ODD updates: One of the artifacts generated by simulation-based testing are simulation traces. These traces contain data such as the position, velocity, and acceleration of the ego vehicle as well those of the other actors over time. The CE loop uses the generated traces to update the requirements and ODD via the specification mining technique, e.g., by applying HyTeM tool [BAR-2023].

## 6. Conclusion and Outlook

In this paper we presented a simulation-based development platform that is designed to verify and validate safety-critical learning-enabled systems in continuous engineering loops. Our findings indicate that in order to be able to deploy machine learning technology in safety-critical automative applications, a careful design of the AI component considering its training data quality, robustness, etc. is necessary, but not sufficient. Before deployment, during design-time, it is essential to analyze whether ODD definition, requirements, and test scenarios give enough coverage to assure the system safety. Therefore, it is needed to have design-time and requirements monitors. Moreover, especially for the vehicles operating in dynamic environments autonomously, run-time monitoring mechanisms are alsoneeded to detect abnormalities and trigger necessary safe actions. In our work, we created an AVP use case as a proof of concept and integrated all such monitors together at different architectural layers. We are convinced that the success of safe deployment of future automated driving systems will depend on such multi-layered design, monitoring and continuous engineering approaches.

During this work, we also paid a special attention to rapidly developing AD and AI safety related standards. Therefore, in the future, we would like to accelerate our efforts to understand how such platforms support the compliance with standards such as ISO/CD TS 5083 - Safety for automated driving systems [Saf-AD], ISO/CD PAS 8800 - Safety and artificial intelligence [Saf-AI], and ISO 21448 - Safety of the intended functionality [SOT-2022].

Another aspect is the modularity of the developed simulation platform. Therefore, standards such as ASAM OpenSCENARIO, ASAM OpenODD and FMI will play a key role. Those standards were partially considered. In our future work, we want to emphasize this modularity, and will also look for appropriate middlewares, such as Siemens PAVE360 [TEM-2022], to achieve seamless data transfer between block using different modelling techniques and languages.

Last not but least, we will have an eye on rapidly developing AI technologies. For example, how generative AI might influence overall system design is a question of interest.

# 7. REFERENCES

[BAR-2023] E. Bartocci, C. Mateis, E. Nesterini and D. Ničković, "Mining Hyperproperties using Temporal Logics," ACM Trans. Embed. Comput. Syst., 2023.

[BEN-2023] S. Bensalem, et al. "Continuous Engineering for Trustworthy Learning-enabled Autonomous Systems", AISOLA workshop, 2023.

[ESE-2020] H. Esen et al. "Validation of Automated Valet Parking". In: Leitner, A., Watzenig, D., Ibanez-Guzman, J. (eds) Validation and Verification of Automated Systems. Springer, Cham. https://doi.org/10.1007/978-3-030-14628-3_16 (2020)

[GRA-2023] R. Gräfe et al., "Large-Scale Application of Fault Injection into PyTorch Models-an Extension to PyTorchFI for Validation Efficiency," in DSN-S, 2023.

[HAV-2018] K. Havelund, D. Peled and D. Ulus. "DejaVu: a monitoring tool for first-order temporal logic," in IEEE Workshop on Monitoring and Testing of Cyber-Physical Systems, 2018.

[LIA-2023] B. H. -C. Liao, C.-H. Cheng, H. Esen and A. Knoll. "Improving the safety of 3D object detectors in autonomous driving using IoGT and distance measures". CoRR abs/2209.10368, (2023).

[MOK-2022] K. Mokos, T. Nestoridis, P. Katsaros and N. Bassiliades, "Semantic Modeling and Analysis of Natural Language System Requirements," in *IEEE Access*, 2022.

[NIC-2020] D. Nickovic and T. Yamaguchi. "RTAMT: online robustness monitors from STL," Proc. of ATVA, 2020.

[PRA-2021] S. Pranger, B. Könighofer, L. Posch, R. Bloem. „TEMPEST - Synthesis Tool for Reactive Systems and Shields in Probabilistic Environments". Automated Technology for Verification and Analysis (2021)

[SAE J3016 - 2021] Taxonomy and Definitions for Terms Related to Driving Automation Systems for On-Road Motor Vehicles. Document J3016_202104. SAE International. doi: https://doi.org/10.4271/J3016_202104

[Saf-AD] ISO/CD TS 5083. "Road vehicles — Safety for automated driving systems — Design, verification and validation" (https://www.iso.org/standard/81920.html), under development

[Saf-AI] ISO/CD PAS 8800. "Road vehicles — "Safety and artificial intelligence" (https://www.iso.org/standard/83303.html), under development

[SOR-2023] L. Sorokin et al. "OpenSBT: A modular framework for search based testing of automated driving systems", CoRR abs/2306.10296, (2023)

[SOT-2022] ISO 21448:2022, "Road vehicles — Safety of the intended functionality" (https://www.iso.org/standard/77490.html)

[TEM-2022] A. Temperekidis, N. Kekatos, P. Katsaros, W. He, S. Bensalem, H. AbdElSabour, M. AbdElsalam, and A. Salem. "Towards a Digital Twin Architecture with Formal Analysis Capabilities for Learning-Enabled Autonomous Systems". MESAS NATO conference for modelling and simulation of autonomous systems, 2022.

[VOO-2023] K. Voogd, J. P. Allamaa, J. Alonso-Mora, T. D. Son. "Reinforcement learning from simulation to real world autonomous driving using digital twin". IFAC (2023)

[WAC-2016] W. Wachenfeld and H. Winner. "The Release of Autonomous Vehicles," in Autonomous Driving: Technical, Legal and Social Aspects, H. Winner, M. Maurer, J. C. Gerdes, and B. Lenz, Eds., Berlin, Heidelberg: Springer, 2016, pp. 425–449.

[WAN-2021] T. Wang, X. Zhu, J. Pang and D. Lin. „FCOS3D: Fully Convolutional One-Stage Monocular 3D Object Detection". https://arxiv.org/abs/2104.10956 , 2021.

[YAN-2021] G. Yang and D. Ramanan, "Learning to Segment Rigid Motions from Two Frames," in *CVPR*, 2021.

[ZHA-2021] X. Zhao, et al. „Assessing the reliability of deep learning classifiers through robustness evaluation and operational profiles". AI Safety, 2021